\documentclass[aps,prl,preprint,groupedaddress]{revtex4}
\usepackage{amssymb,amsmath}

\usepackage{graphicx}
\usepackage{dcolumn}
\usepackage{bm }

\draft

\begin{document}

\title{Resistance
distribution in the hopping percolation model}

\author{Yakov M. Strelniker,
 Shlomo Havlin,  Richard Berkovits,
and  Aviad  Frydman
}

\affiliation{Minerva Center, Jack and Pearl Resnick Institute of
Advanced Technology, and Department of Physics, Bar-Ilan
University, 52900 Ramat-Gan, Israel}

\date{\today}

\begin{abstract}
We study the distribution function, $P(\rho)$, of the effective resistance,
$\rho$, in two and three-dimensional random resistor network of
linear size $L$ in the hopping percolation model.
In this model each bond has a conductivity taken from
an exponential form
$\sigma \propto \exp(-\kappa r)$, where $\kappa$  
 is a  measure of disorder,    and  $r$
is a random number, $0\le r \le 1$.
We find that in 
both
the {\em usual strong disorder} regime 
$L/\kappa^{\nu} > 1$   (not sensitive to removal of
any single bond) and the {\em extreme  disorder} regime
$L/\kappa^{\nu} < 1$  (very sensitive to such a
removal) the distribution depends only on $L/\kappa^{\nu}$ and 
 can be well approximated
by a log-normal function with dispersion $b \kappa^\nu /L$,
where $b$ is a coefficient
which depends on the type of the lattice.
\end{abstract}

\pacs{PACS numbers:
 64.60.Ak, 02.50.-r,
73.23.-b; 72.80.Tm, 78.66.Sq, 73.50.Jt, 77.84.Lf
}

\maketitle

\section{Introduction}
\label{Introduction}

The concepts and methods of percolation theory are widely used
to explain many phenomena in physics, classical
as well as quantum (for review see e.g.
Refs.\  \cite{ShklEfr,HavlinBook,Ahar}).
The canonical model for studying transport properties of disordered
systems is percolation on a lattice. Usually it is also assumed
that the conductivity between neighboring  lattice sites may be defined
as either finite, or zero (i.e., either conducting or insulating)
without loss of generality. This model [which we denote as the
bond (or site) percolation] has been extensively
studied and is understood quite well. For the description
of the nearest neighbor hopping 
in granular materials it is much more natural to define the
conductivity between two  neighboring lattice sites (labeled as
``i'' and ``j'') by $\sigma_{ij} \propto \exp[-r_{ij}/r_0-\epsilon_{ij}/k_BT]$,
 where $r_{ij}$ is the distance between the two sites,  $r_0$
is the scale over which the wave-function outside the grain decays,
$\epsilon_{ij}$ is the energy difference between grains, and $T$
is the temperature. Here  we neglect the thermal hopping term
 (high temperature regime)
and consider only nearest neighbor hopping.
This behavior may be captured by a lattice 
 model for which \cite{StrelBerk,Halperin,Tyc,Le,Sar}
\begin{equation}
\sigma_{ij}=\sigma_0 \exp[-\kappa r(ij)],
\label{hopping}
\end{equation}
where $\kappa$ is a measure of disorder, $r(ij)$ is a random
number taken from uniform distribution in the range
(0,1), and $\sigma_0$ is a dimension coefficient \cite{Halperin,StrelBerk}.
We shall name this model the hopping percolation model.

One might expect that such small differences in the formulation
of the disorder in the $\sigma_{ij}$ (namely,
$\sigma_{ij} = 0,1$ or $\sigma_{ij}=\sigma_0 \exp[-\kappa r(ij)]$) will
lead to no important difference in the global conductance properties
of these systems. Quite surprisingly, recent experiments on the
conductance of granular material \cite{Cohen[1]} might
suggest otherwise. Specifically, the number of red bonds
(which are critical for current) expected
in framework of the traditional percolation theory is proportional
to $L^{1/\nu}$, where $\nu$ is the percolation correlation critical exponent
and $L$ is the system size \cite{Con}. Thus, for a typical experimental set-up
of $10^9$  grains, one expects $10^2$ red bonds. On the other
hand, measurements of transport through Ni granular ferromagnets,
indicates a much lower number of red bonds (typically of order one)
\cite{Cohen[1]}. In a recent paper \cite{StrelBerk} 
we have attributed this difference to the fact that
the estimation of the number of red bonds ($L^{1/\nu}$) is based on the
bond percolation model, while for the hopping percolation we expect
a transition to a regime of extreme strong disorder in which a {\it single}
red bond governs the behavior of the system. The onset of this regime
scales as to $\kappa/L^{1/\nu}$.

It is important to note that in contrast to the traditional bond
(or site) percolation model, in which the system is either a metal
or an insulator, for the  hopping percolation model the system
always conducts some current.
Hopping conductivity (i.e., exponential local resistance)
(\ref{hopping}) is always associated with {\em strong disorder}. As
was shown in Ref. \cite{StrelBerk}, there are two regimes within
this strong disorder: a regime which is not sensitive to the removal
of a single bond, as expected from the usual percolation theory,
termed the {\em usual strong disorder} regime \cite{rem}. 
While for even stronger disorder a regime which is very sensitive to the
removal of a specific single bond exists, denote as {\em extreme
disorder}.
 In the extreme 
 disorder regime
 a single bond can
determine the transport properties of the entire macroscopic system
\cite{StrelBerk,Cohen[1]}.

The remainder of this paper is arranged as follows.
In Section \ref{Model}
we describe our model and the  numerical approach.
In Section \ref{Results}
we present some numerical results, followed by a brief discussion
in Section \ref{Summary}.

\section{Model}
\label{Model}

\begin{figure}
\centerline{
\includegraphics[height=8.cm]{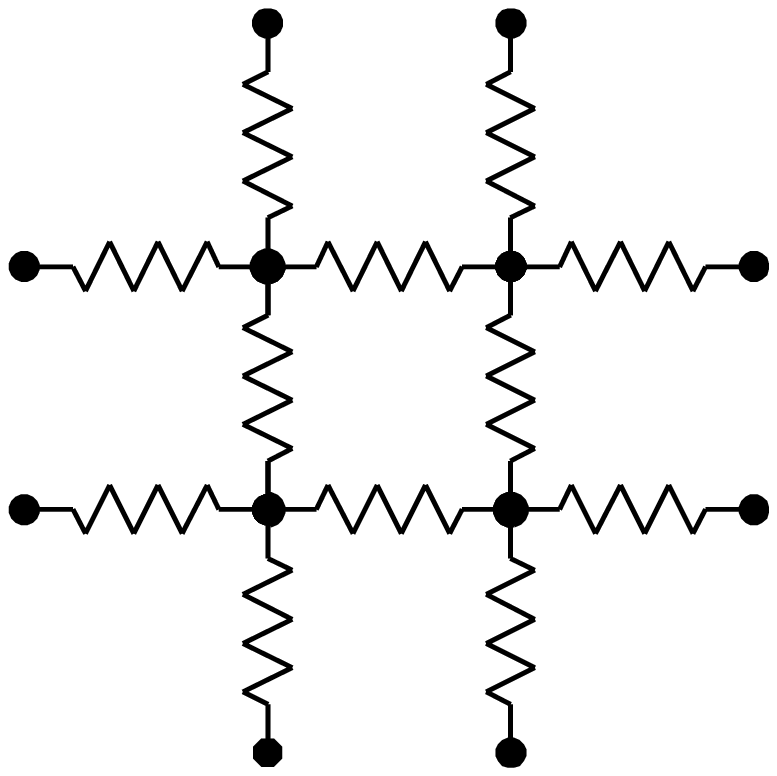}
\hspace{8.cm}
}
\vspace{-8.8cm}
\centerline{
\hspace{4.cm}
\includegraphics[height=6.cm]{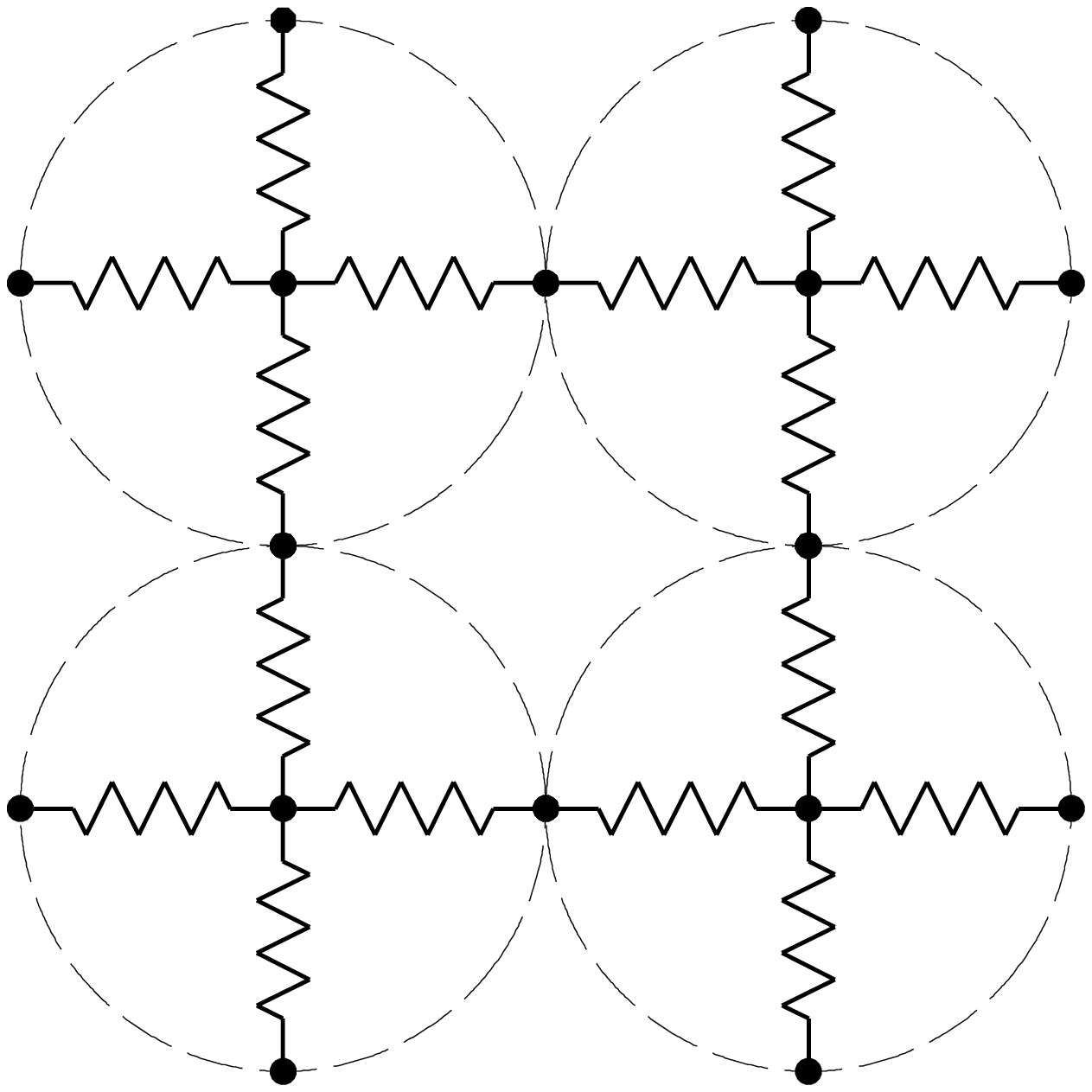}
}
\vspace{-0.5cm}
\centerline{{\large {\bf a.}}~~~~~~~~~~~~~~~~~~~~~~~~~~~~~~~~{\large {\bf b.}}}
\vspace{0.5cm}
\vspace{-0.5cm}
\caption{\label{network} 
(a) A square  bond percolation net of resistors  with random
resistivity given by Eq.\ (\protect\ref{hopping}), where 
$(ij)$ denotes the bond between sites $i$ and $j$.
(b) A site percolating network.
The resistivity of all four resistors
within a dashed circle is determined by a single random number
$r(ij)$, where $ij$ denotes the labeling of the grid point $i,j$.
}
\end{figure}

We perform  large-scale Monte Carlo simulations
for  calculating transport in these
systems.
We build a bond-percolating
 Miller-Abrahams like resistor
network (see Fig.\ \ref{network} and
 Ref.\ \onlinecite{StrelBerk,MillerAbrahams,Kirkpatrick,Sarychev2}),
but assume the conductivity of each resistor
to have the
form  given in Eq.\ (\ref{hopping}).
Then solve the  corresponding set
 of linear Kirchhoff equations and calculate the total effective
resistance $\rho_e$ for two dimensional (2D)  and three dimensional
(3D) networks
\cite{StrelBerk,MillerAbrahams,Kirkpatrick,Sarychev2}.

\begin{figure}
\centerline{
\includegraphics[height=7.cm]{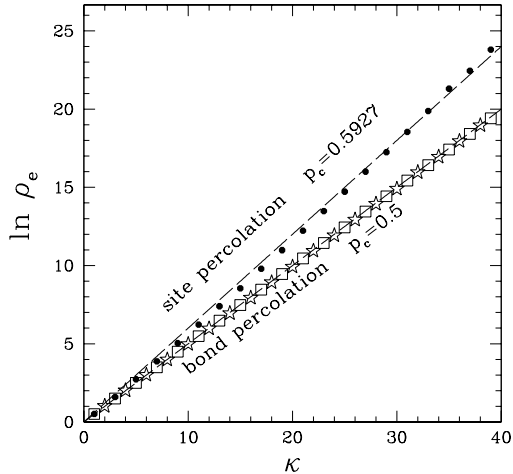}
}
\caption{\label{NoCut}
A semi-log plot of the averaged resistance
$\rho_e$ vs. $\kappa$.
In the case of site percolation
($p_c=0.592746$) the slope
of the curve is close to $0.6$, while for the case of
bond percolation ($p_c=0.5$)
this slope is equal to 0.5
[cf. with  Eq.\ (\protect\ref{exact})].
The system sizes shown are: 
$L=20$ (solid circles), $40$ (open squeres),
$100$ (open stars).
}
\end{figure}

We begin by calculating
 the average effective conductivity $\sigma_e$.
 The approximate expression
 \cite{Halperin,Ahar,ShklEfr} for the effective conductivity
 $\sigma_e$, of  a
 random resistor network \cite{Kirkpatrick} with
 local conductivities
given by
Eq.\ (\ref{hopping}), in
2D
is
\begin{equation}
 \sigma_e = \sigma_{0} e^{-p_c \kappa}.
\label{exact}
\end{equation}
In Fig.\ \ref{NoCut} we show this dependence (in terms
of resistivity $\rho_e=1/\sigma_e$) for both site and bond percolations
(see Fig. \ref{network}) for different lattice sizes.
In Ref.\ \cite{Tyc} it was shown that
in the limit $\kappa \rightarrow \infty$,
Eq.\ (\ref{exact})
 is
 exact.
It is easy to show
 that in the case of  2D  random resistor bond network
(for which $p_c=0.5$),
Eq.\ (\ref{exact}) follows immediately from
the  Keller-Dykhne theorem
and
 is
 exact
for arbitrary $\kappa$ \cite{Keller2}. 
Similarly this results can be found
in framework of the symmetric
 self-consistency effective-medium approximation (EMA)
rewritten for many component composite \cite{Landauer}.
 From Eq.\ (\ref{exact}) follows that
the effective conductivity
$\sigma_e$ depends on $\kappa$ and does not
depend on the system size $L$.
For finite $L$, Eq.\ (\ref{exact}) represents the mean conductivity over all
configurations of the disordered system.

 Next we
study the
fluctuations of the resistance, $\rho=1/\sigma$, from the  mean value  $\rho_e=1/\sigma_e$
for individual systems of finite size $L$.
We perform  numerical calculations of the probability
distribution function $P(\rho)$ (i.e., the probability
that the total resistance of the system is
 $\rho$) as well as  the variance
${\rm var}(\rho)$ as a function of $L$ and $\kappa$.
As  shown in Ref.\ \cite{StrelBerk}, the
relative variance 
(in contrast to $\rho_e$),
strongly  depends on
$L$ and $\kappa$  {\it only}
through the
scaled variable
 $h \equiv L/\kappa^{\nu}$.
 Here $\nu$ is
 the  critical exponent of  the  percolation
 correlation
length $\xi \propto (p-p_c)^{-\nu}$
(in  2D $\nu=4/3\simeq1.33$, while in
3D $\nu\simeq0.88$
\cite{ShklEfr,HavlinBook,Ahar}).
It was also shown \cite{StrelBerk} that
$h$ describes the transition from
strong disorder
($h>1$) to
extreme disorder
($h<1$).

\begin{figure}
\centerline{
\includegraphics[height=9.cm]{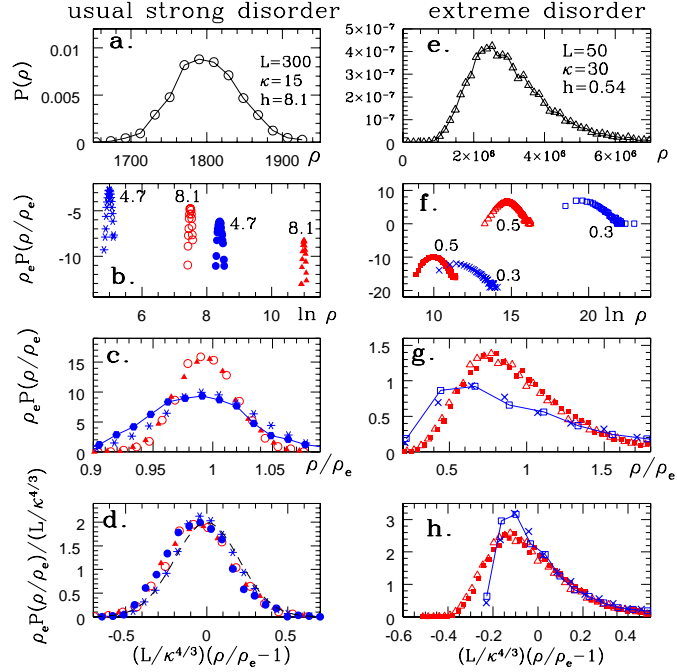}
}
\caption{\label{StrongDis}
(a)-(d) The probability distribution for  the case of
usual strong
disorder ($h>1$).
(a) A typical form of
$P(\rho)$ vs. $\rho$ for
$h=8.1$.
(b)
Similar to
(a), but shown in log-log scale
for
$L=300$, $\kappa=15$, $h=8.1$ ($\circ$),
 $L=500$, $\kappa=22$, $h=8.1$
($\blacktriangle$),
$L=100$, $\kappa=10$, $h=4.7$ ($\ast$),  and
$L=200$, $\kappa=16.84$, $h=4.7$ ($\bullet$).
 (c) Scaling plots $\rho_eP(\rho/\rho_e)$ vs.
$\rho/\rho_e$ for  the results shown in
(b).
Only the data with the same value of $h$
scale to a  unique function.
 (d) Scaling plots of
 $\rho_e P(\rho/\rho_e)/(L/\kappa^\nu)$ vs.
 $(L/\kappa^{\nu})(\rho/\rho_e-1)$
lead to the  collapse of the data
 shown in
(b) and (c).
The dashed line in
 (d) is the
Gaussian  distribution (\protect\ref{Gaus})
 with $b=0.2$.
(e)-(h) Similar to
 (a)-(d), but for  systems
with extreme
 disorder ($h<1$):
 $L=50$, $\kappa=30$, $h=0.54$
($\triangle$),
 $L=30$, $\kappa=20.4$,   $h=0.54$
($\blacksquare$),
$L=40$, $\kappa=40$,  $h=0.3$
($\square$) and $L=20$, $\kappa=23.78$,  $h=0.3$
($\times$).
(h)
Similar process like
(d),
 does not
lead to collapse  in
 the case of
extreme
disorder.
The lines connecting the points for $h=4.7$ in
 (c)  and for $h=0.3$ in
 (g) and (h)
are guides to the eye.
}
\end{figure}

\section{Results}
\label{Results}

\begin{figure}
\centerline{
\includegraphics[height=8.cm]{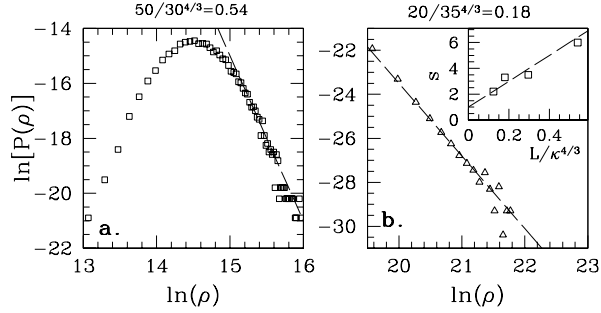}}
\vspace{-4.cm}
\caption{\label{LogNormPowerLaw}
A log-log plot of
 $P(\rho)$ vs. $\rho$.
By increasing disorder the log-normal distribution
transforms to the power-law $\rho^{-s}$.
(a) $L=50$, $\kappa=30$
($h=0.54$).
(b) $L=20$, $\kappa=35$
($h=0.18$).
Inset:
The exponent  $s$ (of the power law $\rho^{-s}$)  vs.
$h$.
By increasing disorder (i.e., decreasing
 $h$)
the exponent $s$ tends to   1.
 $L=50$, $\kappa=30$
($h=0.54$);
$L=14$, $\kappa=18.19$
 ($h=0.3$);
$L=20$, $\kappa=35$
($h=0.18$);
 $L=14$, $\kappa=35$
($h=0.12$).
$\nu=4/3$.
}
\end{figure}

Here we present  numerical results suggesting that
$P(\rho)$ also depends {\it only} on
$L/\kappa^\nu$.
In order to verify and to quantify this hypothesis, we
study numerically
$P(\rho)$ for systems of
different sizes $L$ and  different
disorder $\kappa$,
but with
the same value of $h$ (see also Ref.\ \cite{Wu}).
In Figs.\  \ref{StrongDis} and \ref{LogNormPowerLaw}
we show
 $P(\rho)$ vs.
$\rho$ for the cases of
strong  and extreme
disorder.
All data corresponding to the same parameter $h$ scale according
 to the same law
[see Figs.\ \ref{StrongDis}(c) and \ref{StrongDis}(g)].
Thus, our  results suggest that $P(\rho)$ is a
 function of both $\rho/\rho_e$ and $h$, i.e.,
\begin{equation}
P(\rho)=\frac{1}{\rho_e} f\left(\frac{\rho}{\rho_e};h\right).
\label{Scaling3}
\end{equation}
Here $h$ determines the form of the  function
and  $P(\rho)$ for a fixed $h$ depends only on $\rho/\rho_e$.
Fig.\ \ref{StrongDis}(c) suggests that in the
strong
 disorder, $h$
controls the  width or standard deviation
 of the rescaled distribution.
Since the standard deviation
 increases when $h$ decreases, we assume
that the standard deviation
$\delta\simeq b \rho_e/h=
b \rho_e \kappa^{\nu}/L$,
where  $b$ is a
parameter which depends on the type of  lattice.
 Indeed, when we plot
 in Fig.\ \ref{StrongDis}(d) $P(\rho/\rho_e) \delta$
 vs. $(\rho-\rho_e)/\delta$,
a collapse of the two plots shown
in Fig.\ \ref{StrongDis}(c) is obtained.
The functional form obtained in Fig.\ \ref{StrongDis}(d) suggests
 that the probability distribution
can be approximated by a Gaussian
\begin{equation}
P(\rho)\simeq
(\sqrt{2\pi} \delta )^{-1}
\exp \left[-
\left(\rho-\rho_e \right)^2 /
2 \delta^2 \right].
\label{Gaus}
\end{equation}
Indeed, the dashed line in Fig.\ \ref{StrongDis}(d) represents a good
 fit to the Gaussian given  by  Eq.\ (\ref{Gaus}).
However, Eq.\ (\ref{Gaus})
can not approximate the asymmetric form of
$P(\rho)$ at
extreme
 disorder
 [see Figs.\ \ref{StrongDis}(g) and \ref{StrongDis}(h)].
We suggest, as will be justified below,
that
$P(\rho)$
can be approximated (in all regimes of disorder) by
the
 log-normal form
\begin{equation}
P(\rho)\simeq
\frac{1}{\sqrt{2\pi} \mu \rho}
\exp \left[
-\frac{ \ln^2 ( \rho/ \rho_e )}
{2 \mu^2}
 \right],
\label{LogNorm}
\end{equation}
where
  $\mu=\delta/\rho_e=b\kappa^\nu/L$.
In fact, Eq.\ (\ref{LogNorm}) includes also the
usual strong
disorder case,
 since
in  the latter case
  $\ln^2(\rho/\rho_e)\simeq (\rho/\rho_e-1)^2$
and Eq.\ (\ref{LogNorm})  reduces to the
  Gaussian form
 (\ref{Gaus}),
while
 at
extreme
disorder  ($\mu \gg 1$) the exponent function
in Eq.\ (\ref{LogNorm}) tends to 1, and
$P(\rho)$
 transforms to the power-like
dependence
$\sim 1/\rho$ (see Fig.\ \ref{LogNormPowerLaw}).

From Eq.\ (\ref{LogNorm}) it follows that
 at $\mu \rightarrow 0$, the distribution function $P(\rho)$
reduces to a delta-function:
 $\lim\limits_{\mu  \rightarrow 0} P(\rho)
=\frac{1}{\rho}
 \lim\limits_{\mu \rightarrow 0} \frac{1}{\sqrt{2 \pi} \mu}
e^{-\frac{\ln(\rho/\rho_e)^2}{2 \mu^2}}=
\frac{1}{\rho}\delta(\ln \rho-\ln \rho_e)=\delta(\rho-\rho_e)$.
Therefore, at $\mu \rightarrow 0$ (i.e.,  $\kappa \rightarrow 0$
or $L  \rightarrow \infty$) the total resistivity of the system
is exactly
 $\rho_0 e^{p_c \kappa}$
 and has no size dependence:
$\lim\limits_{\mu \rightarrow 0} \int \rho P(\rho) d \rho=
\rho_e$.

It should be noted that a log-normal distribution of resistances is found 
in quantum models of hopping conductivities 
 (see e.g., Ref.\
\cite{Shapiro} and references therein), while
here it is demonstrated for classical exponential disorder
(\ref{hopping}).
Moreover, our  result (\ref{LogNorm}) yield the specific analytical 
form of $P(\rho)$, which includes the dependence on $\kappa$ and $L$
for all regimes of disorder.

\begin{figure}
\centerline{
\includegraphics[height=9.cm]{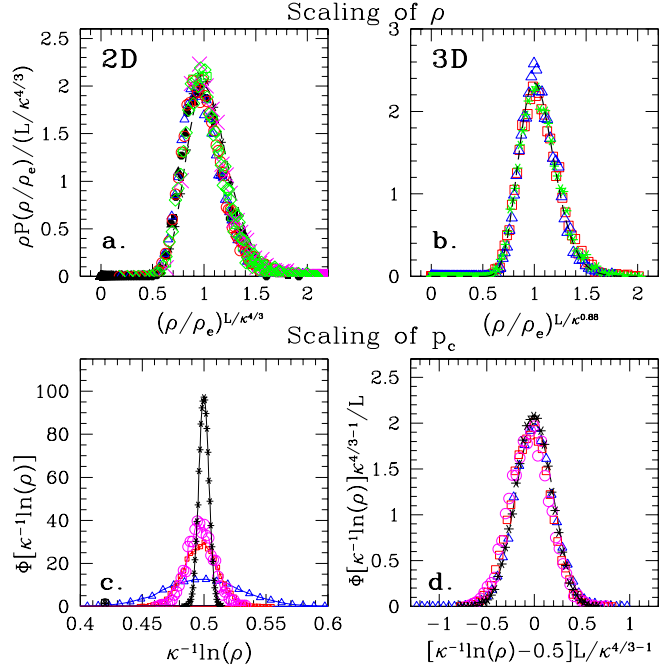}
}
\caption{\label{LogNorm2D}
(a) A scaling  plot of
 $\rho P(\rho/\rho_e )$ vs. $(\rho/\rho_e )^{L/\kappa^\nu}$
of the data plotted in
Fig.\  \ref{StrongDis} (2D case).
The dashed line represents  the analytical result
(\protect\ref{LogNorm}),
with $b=0.2$ and
 $\nu=4/3$ for 9 systems with $L=500$, $\kappa=22$
 ($\blacktriangle$); $L=300$, $\kappa=15$
 ($\blacksquare$);  $L=200$, $\kappa=10$
 ($\bullet$);  $L=100$, $\kappa=10$
($\ast$);  $L=50$, $\kappa=30$
($\triangle$); $L=40$, $\kappa=40$ ($\square$);
  $L=30$, $\kappa=20$
($\circ$);
  $L=20$, $\kappa=40$ ($\times$);
  $L=20$, $\kappa=30$ ($\lozenge$);
(b) Similar to
(a),
but for 3D, with $b=0.18$ and $\nu=0.88$
for three systems with
$L=20$, $\kappa=15$ ($\square$);
 $L=26$, $\kappa=15$ ($\triangle$);
 $L=10$, $\kappa=6.8$ ($\ast$).
($h=1.85$).
The dashed line is the  analytical result, Eq.\
 (\protect\ref{LogNorm}).
(c) $\Phi(p_c)= \Phi[\ln(\rho)/\kappa]$ vs. $ p_c$,
where $ p_c= \kappa^{-1}\ln(\rho)$
for different values of
the ratios $L/\kappa^{\nu}$:
$L=100$, $\kappa=10$, $h=4.68$  ($\ast$);
$L=60$, $\kappa=30$, $h=0.65$ ($\circ$);
$L=40$, $\kappa=20$, $h=0.74$ ($\square$);
$L=20$, $\kappa=30$, $h=0.22$ ($\triangle$);
(d) Scaling
 $\Phi[\ln(\rho)/\kappa]/(L/\kappa^{\nu-1})$ vs.
 $[\ln(\rho)/\kappa-0.5](L/\kappa^{\nu-1})$ with
$\nu=4/3$.
The  values of the ratios $L/\kappa^{\nu}$ are the
same as in (c).
The dashed line is the analytical fit,
as derived from Eq.\ (\protect\ref{LogNorm}).
}
\end{figure}

In Fig.\ \ref{LogNorm2D}(a)  we test Eq.\ (\ref{LogNorm}) by comparing
it to simulation results.
It is shown that the numerical results
of the 2D resistance $\rho P(\rho/\rho_e)/(L/\kappa^\nu)$
scale vs. $(\rho/\rho_e)^{L/\kappa^\nu}$, as
predicted by Eq.\ (\ref{LogNorm}) for both
strong and extreme disorder.
A similar plot is presented in Fig.\ \ref{LogNorm2D}(b)
for a 3D lattice.
Although for the 3D case Eq.\ (\ref{exact}) is not exact
(since Keller-Dykhne theorem exists only in 2D),
nevertheless the approximated expression
$\sigma_e \simeq  \sigma_0\kappa^\nu e^{-p_c \kappa}$
is known \cite{Halperin,Tyc}, resulting in the distribution law
(\ref{LogNorm}). Since in 
Eq. (\ref{exact})
the parameter $\kappa$ appears with the prefactor $p_c$,
we should expect that $p_c$ enters into the parameter
$\mu$ of Eq.\ (\ref{LogNorm}) as $\mu=\alpha (p_c \kappa)^\nu/ L$. 
Comparing the values $b=0.2$ observed for the square bond percolation lattice
($p_c=0.5$)
and $b=0.18$ for the cubic site percolation ($p_c=0.3116$), 
we find that $\alpha=0.503$. The dependence of $\mu$ on
$p_c$ is in agreement with result of Ref.\ \cite{Kalis}.
These results  strongly support our proposition that Eq.\ (\ref{LogNorm})
describes well the distribution in all ranges of disorder.

\begin{figure}
\centerline{
\includegraphics[height=9.cm]{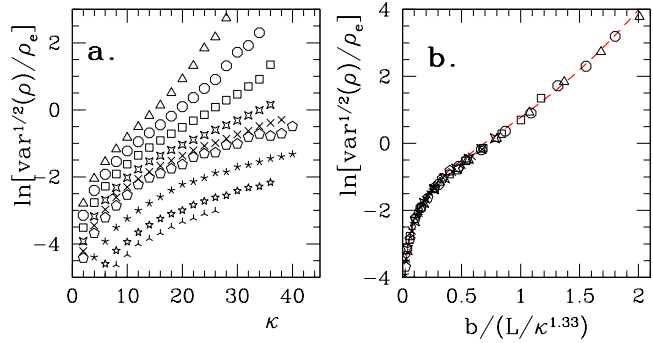}
}
\vspace{-4.2cm}
\caption{\label{StrongDisScaling}
(a) A semi-log plot of the
relative
variance $[{\rm var}(\rho)]^{1/2}/\rho_e$ vs. $\kappa$
for various sizes of the system:
$L=10, 14,20,30,40,50, 100,200,300$ (from top to bottom).
(b) A semi-log scaling plot of the same quantity
 vs. $\mu=b\kappa^{\nu}/L$, where $b=0.2$, $\nu=4/3$.
The dashed line represents
Eq.\ (\protect\ref{Var}).
}
\end{figure}

The  variance ${\rm var}(\rho)$
can be expressed as
$
\langle
\rho^2 \rangle-\rho_e^2
$, where
$\langle \rho^{n} \rangle =\int_{\rho_{\rm min}}^{\rho_{\rm max}}
 \rho^n P(\rho) d \rho$, and
$\rho_e=\langle \rho \rangle$.
For large enough
 $\kappa$,
$
\langle \rho^{n} \rangle=\int_0^{\infty} \rho^{n}
P(\rho)d \rho=
\rho_0 e^{n \ln \rho_e+\frac{1}{2}n^2 \mu^2}
$
 and
the relative variance
takes the form
\begin{equation}
\left[{\rm var}(\rho)\right]^{1/2}
/\rho_e =
[e^{\mu^2}(e^{\mu^2}-1)]^{1/2}
\label{Var}
\end{equation}
 (see Ref.\ \cite{Aitchison}).
Fig.\ \ref{StrongDisScaling} presents numerical results
showing that the relative variance scales
as a function of $\mu=b\kappa^\nu/L$ in accordance with Eq.\ (\ref{Var}).

Next we shall present analytical arguments for the log-normal  distribution
(\ref{LogNorm}).
According to the central limit theorem \cite{Aitchison},
if the values of $\ln \rho$ are
normally distributed,
then the values of  $ \rho$ should follow the log-normal distribution.
Assuming  $\ln \rho=\kappa p_c$
[see Eq.\ (\ref{exact})] for all $\rho$, the distribution
$P(\kappa^{-1}\ln \rho)$ is simply 
the distribution of the percolation threshold
$\Phi(p_c)$ which is normally distributed (e.g., Refs.\
\cite{Lev}).
Indeed, in Fig.\ \ref{LogNorm2D}(c) we show that $ \Phi[\kappa^{-1}\ln(\rho)]$
approximately follows a normal distribution centered at $p_c=0.5$.
Thus, the distribution of $\rho$ should be log-normal as in
 Eq.\ (\ref{LogNorm}). 

Using the above assumption $\ln \rho =\mu/\kappa=\kappa p_c$, it is possible to evaluate 
the distribution $\Phi(p_c)$ and its
standard deviation $\delta_{p_c}$. 
One can write a simple relation $\Phi(p_c)dp_c=\Phi^{\prime}(y)dy$,
where $y=\kappa p_c$,
and get
$\Phi(p_c)=\Phi^{\prime}(y)\frac{dy}{dp_c} =\kappa \Phi(y)$.
Therefore, $\Phi(p_c)=\kappa \Phi^{\prime}(\ln \rho)
\sim \kappa \mu^{-1}
\exp[-\kappa^2(p_c- \bar p_c)^2/2\mu^2]$
$=\delta_{p_c}^{-1}
\exp[-(p_c- \bar p_c)^2/2\delta_{p_c}^2]$, where $\bar p_c$ is the
mean value of the percolation threshold and
 $\delta_{p_c}=\mu/\kappa$ is the standard deviation of $\Phi(p_c)$.
This form of $\Phi(p_c)$ is supported by our numerical
simulations shown in Figs.\ \ref{LogNorm2D}(c)
and \ref{LogNorm2D}(d).
 
This specific form for $\delta_{p_c}=b\kappa^{\nu-1}/L$
in the hopping percolation model should
be compared to $\delta_{p_c} = L^{-1/\nu}$ known for the bond percolation model
\cite{Lev,Con}.
This further emphasizes the
 differences between the bond percolation
model considered in Ref.\ \cite{Lev} and  the
hopping percolation model considered here.

\section{Summary}
\label{Summary}

In summary,  we find the specific form of the resistance distribution in the hopping percolation model.
For all ranges of strong disorder $\kappa$ and lattice sizes  $L$,
 the  distribution is log-normal and depends only  on the ratio
 $\kappa^{\nu}/L$,
where $\nu$ is the correlation exponent for the bond percolation case.
Assuming the relation $\rho=\exp(\kappa p_c)$ for finite systems leads to a variance of $p_c$,
 $\delta_{p_c}=\kappa^{\nu-1}/L$, which is different from
 $\delta_{p_c}=L^{-1/\nu}$ known for the bond percolation model \cite{Lev}.
Our results may be relevant to ac conductivity measurements in such systems.
By appropriate choice of frequency one can detect regions of  size 
smaller than $\kappa^\nu$,
where a crossover in behavior from extreme to usual strong disorder behavior is expected.

\begin{acknowledgments}

This research was supported in part by grants from the
US-Israel Binational Science Foundation, the Israel Science Foundation,
and the KAMEA Fellowship  program of the
Ministry of Absorption  of the State of Israel.

\end{acknowledgments}

\end{document}